\documentclass[twocolumn,aps,preprintnumbers,amsmath,amssymb,nofootinbib]{revtex4}

\usepackage{graphicx}
\usepackage{color}
\usepackage{soul}
\usepackage{latexsym}
\usepackage{epsf}
\usepackage{amsmath}

\def\lsim{\mathrel{\rlap{\lower4pt\hbox{\hskip1pt$\sim$}}
    \raise1pt\hbox{$<$}}}         
\def\gsim{\mathrel{\rlap{\lower4pt\hbox{\hskip1pt$\sim$}}
    \raise1pt\hbox{$>$}}}
\def\fun#1#2{\lower3.6pt\vbox{\baselineskip0pt\lineskip.9pt
  \ialign{$\mathsurround=0pt#1\hfil##\hfil$\crcr#2\crcr\sim\crcr}}}

\input epsf

\def\beq{\begin{equation}}
\def\eeq{\end{equation}}
\def\bea{\begin{eqnarray}}
\def\eea{\end{eqnarray}}

\def\ln#1{\mathrm{ln}\left(#1\right)}

\newcommand{\degree}{\ensuremath{^\circ}}

\newcommand{\mN}{m_{\scriptscriptstyle N}}
\newcommand{\mX}{m_{\scriptscriptstyle \chi}}
\newcommand{\rhoX}{\rho_{\scriptscriptstyle \chi}}

\newcommand{\vsun}{v_{\odot}}

\begin{document}

\preprint{IPMU-10-0002}
\title{Neutrino Constraints on Inelastic Dark Matter after CDMS II}
\author{Jing Shu~$^{\bf a}$}
\email{jing.shu@ipmu.jp}
\author{Peng-fei Yin~$^{\bf b}$}
\email{s7dx5v3@pku.edu.cn}
\author{Shou-hua Zhu~$^{\bf b}$}
\email{aiwen_fan@pku.edu.cn} \affiliation{
$^{\bf a}$Institute for the Physics and Mathematics of the Universe, The University of Tokyo, Chiba $277-8568$, Japan\\
$^{\bf b}$Institute of Theoretical Physics $\&$ State Key Laboratory of Nuclear Physics and Technology, Peking University, Beijing 100871, P.R. China \\
}

\begin{abstract}

We discuss the neutrino constraints from solar and terrestrial dark
matter (DM) annihilations in the inelastic dark matter (iDM)
scenario after the recent CDMS II results. To reconcile the
DAMA/LIBRA data with constraints from all other direct experiments,
the iDM needs to be light ($m_\chi < 100$ GeV) and have a large
DM-nucleon cross section ($\sigma_n \sim$ 10$^{-4}$ pb in the
spin-independent (SI) scattering and $\sigma_n \sim$ 10 pb in the
spin-dependent (SD) scattering). The dominant contribution to the
iDM capture in the Sun is from scattering off Fe/Al in the SI/SD
case. Current bounds from Super-Kamiokande exclude the hard DM annihilation channels, such as
$W^+W^-$, $ZZ$, $t\bar{t}$ and $\tau^+ \tau^-$. For soft channels
such as $b\bar{b}$ and $c \bar{c}$, the limits are loose, but could
be tested or further constrained by future IceCube plus DeepCore.
For neutrino constraints from the DM annihilation in the Earth, due
to the weaker gravitational effect of the Earth and inelastic
capture condition, the constraint exists only for small mass
splitting $\delta <$ 40 keV and $m_\chi \sim (10, 50)$ GeV even in
the $\tau^+ \tau^-$ channel.

\end{abstract}
\maketitle

\section{Introduction}

A wide variety of cosmological observations, which include the
highly precise measurements of the cosmic microwave background
\cite{Komatsu:2008hk}, galactic rotation curves
\cite{Borriello:2000rv}, the weak gravitational lensing of distant
galaxies by foreground structure \cite{Hoekstra:2002nf}, and the
weak modulation of strong lensing around individual massive
elliptical galaxies \cite{Metcalf:2003sz}, all indicate that about
22$\%$ of our universe consists of non-baryonic, non-luminous dark
matter (DM). The nature of DM, on the other hand, still remains a mystery.

The first direct detection experiment that provided strong
evidence for dark matter is the DAMA collaboration, which announced
an 8.3 $\sigma$ discovery \cite{Bernabei:2008yi} in the annual
modulations of nuclear recoil rate. In the simplest DM models, such
an event rate is excluded by two orders of magnitude in other low-background nuclear experiments. A widely invoked resolution of this controversy is the inelastic dark matter (iDM) model
\cite{TuckerSmith:2001hy,Chang:2008gd}, which introduces a
transition to an excited state in the dark matter nucleus
scattering. If the mass splitting $\delta$ is roughly 100 keV, which is close
to the kinetic energy of a DM in the halo, the kinematics is
significantly modified and the controversy can be settled.

Very recently, the Cryogenic Dark Matter Search (CDMS II) announced
the observation of two signal events \footnote{CDMS has a very low
background comparing to the other direct detection experiments and
the two observed events are well separated from other events that
have failed to pass the cuts. In other
experiments, for instance XENON \cite{Angle:2007uj}, many observed
events are at the border of the cuts and are used to set limits on
DM scattering cross section.} with a 77$\%$ confidence level and
additional two events just outside the signal region border
\cite{Ahmed:2009zw}. Statistically this is not significant to claim
a discovery, nevertheless, it is very suggestive if one believe such
signals are coming from DM nuclei scattering. In particular, these two events with small
recoil energies suggest that the
mass of dark matter is smaller than 100 GeV \footnote{Interestingly, the preferred DM mass region largely overlaps with the one used to explain DAMA in the iDM model.} with SI cross section
$\sim$ O(10$^{-8}$) pb which could escape other experimental constraints. Many models have been proposed since then \cite{CDMS}.
In addition, limits on the dark matter
nucleus cross section are obtained from the data. In the inelastic spin-independent (iSI) DM
models, only a very narrow region of parameter space is allowed
\cite{Ahmed:2009zw, Angle:2007uj, SchmidtHoberg:2009gn,Kopp:2009qt}
, but the constraint can be large relaxed if one consider the inelastic
spin-dependent (iSD) DM models \cite{Kopp:2009qt}.

Other experiments that give a strong constraint on the
DM-nucleon cross section besides the direct detection experiments are the neutrino telescopes which are detecting the neutrino flux from DM captured and annihilated in the Sun
or Earth. For example, Ref. \cite{Hooper:2008cf} and Ref.
\cite{Nussinov:2009ft,Menon:2009qj} give the neutrino constraints to
the light elastic DM (eDM) and iSI DM which could account for DAMA results
respectively. When the DM nucleus scattering is iSD, the constraints
by the neutrino flux from the Sun can be dramatically different. The
reason is that the kinetic energy requirement for a DM ($\chi$)
scattering inelastically with a nucleus ($N$) is
\bea \label{eq:
Econdition} E_\chi > \delta \left(1 + \frac{m_\chi}{m_N} \right) \ .
\eea
For light elements like hydrogen, only very energetic DM particles can scatter with them and get
captured in the Sun if $\delta$ is not very small. Then the overall capture rate with their dominant contributions from light elements in the Sun will be significantly reduced. The modified kinematics for
inelastic scattering in Eq. (\ref{eq: Econdition}) also largely
reduced the kinetic phase space of DM captured in the Earth because of the
small escape velocity of the Earth. Therefore, it is of importance to preform a
detailed analysis regarding the above issues motivated by the recent
DAMA, CDMS results.

The paper is organized as follows. In Section \ref{sec: crsun}, we
calculate the capture rate for both iSI and iSD DM in the Sun and
we discuss the possible uncertainties in the calculation. In Section
\ref{sec:limnt}, we obtain the constraint on the DM-nucleon
inelastic cross section for different annihilation channels from Super-Kamiokande (Super-K) and the expected event number from the on-going large volume neutrino telescope, IceCube.
In Section \ref{sec:limearth}, we will discuss the neutrino constraint from iDM
captured in the Earth. Finally we conclude and discuss our studies in Section
\ref{sec:con}.

\section{Capture Rate For Inelastic Dark Matter in the Sun}
\label{sec: crsun}

When DM particles travel through the Sun system, they may get
trapped by the gravitational potential and continue to lose energy
with every collision against nucleus and captured in the Sun. The
time evolution of the DM population in the Sun is given by the
differential equation
\begin{equation}
\label{eq:CArate}
 \dot{N} = C_{\odot} - C_A N^2-C_E N,
\end{equation}
where $C_{\odot}$ is the capture rate of the DMs in the Sun and
annihilation rate is $\Gamma_A = C_A N^2 / 2$ with $C_A= \langle \sigma
v \rangle /V_{\odot eff}$, $C_E$ is the inverse time for DM to escape via
evaporation which will be neglected below since our DM is not very light. For typical DM we have considered, the time scale required to reach
equilibrium between capture and annihilation, $t_{eq} = 1 /
\sqrt{C_{\odot} C_A}$, is much smaller than the age of the solar
system so the annihilation rate is saturated at half of the capture
rate $\Gamma_A = C_{\odot}/2$. Therefore, calculating the DM capture
rate in the Sun is crucial to estimate the final neutrino flux from
the Sun \cite{Nussinov:2009ft,Menon:2009qj, Gould:1987ir,
Jungman:1995df, Wikstrom:2009kw}.

Suppose a DM particle has a velocity $u$ very far away from the Sun,
at some point the escape velocity in a shell is
$v(r)$, the DM velocity $w$ will be
\begin{equation}
w^2=u^2+v(r)^2.
\end{equation}
In the DM-nucleus center of mass frame, it is easy to see that in
order to have a DM-nucleus inelastic scattering, the following
condition is required
\begin{equation}
\frac{\mu w^2}{2}>\delta \label{c1} ,
\end{equation}
where $\mu \equiv m_{\chi} m_N / (m_\chi + m_N)$ is the DM-nucleus
reduced mass, $\delta$ is the mass split between two DM states. The
escape velocity $v(r)$ is approximated as \cite{Gould:1991hx}
\begin{equation}
v^2(r)=v_c^2-\frac{M(r)}{M_{\odot}}(v_c^2-v_s^2),
\end{equation}
where $M_{\odot}$ is the mass of the Sun and $v_c=1354$ km/s and
$v_s=795$ km/s. $M(r)$ is the mass profile of the Sun which can be
obtained by the Sun density profile \cite{Gandhi:1995tf}
\begin{equation}
\label{density} \rho(r)=\rho_0 \exp(-B \frac{r}{R_{\odot}}) \ ,
\end{equation}
where $\rho_0 = 236.93$ g/cm$^3$ and $B= 10.098$.

The DM capture rate in the Sun is given by
\begin{equation}
C_{\odot} = \int 4 \pi r^2 dr \int du \frac{f(u)}{u} w  \Omega(w) \
,
\end{equation}
where $f(u)$ is the DM velocity distribution at infinity,
$\Omega(w)$ is the rate per unit time that a DM with velocity $w$
scatter to a velocity less than $v$. We assume the DM velocity
distribution far from the Sun follows a Maxwell-Boltzmann
distribution in the Sun rest frame
\begin{equation}
 f(x) dx = \frac{\rhoX}{\mX} \frac{4}{\sqrt{\pi}} x^2 e^{-x^2} e^{-\eta^2}\frac{\sinh (2 x\eta)}{2x\eta}dx
\end{equation}
where the dimensionless variables are defined as $x^2 \equiv
u^2/v_0^2$ and  $\eta^2 \equiv \vsun^2/v_0^2$, $v_0=$220 km/s is the
average local circular velocity which is related to DM velocity
dispersion $\bar{v}$ as $v_0^2/\bar{v}^2=2/3$.

Taking into account the form factor suppression, the scattering
function $\Omega (w)$ is given by
\begin{eqnarray}
\Omega(w) &=& \sum_i \frac{n_i w}{Q_{i \; max} - Q_{i \; min}}\int_{Q_{i \; min}^{'}}^{Q_{i \; max}} dQ_i \nonumber \\
& \times & \left(\sigma_i^{SI} F_i^2(Q_i) + \sigma_i^{SD} S_i (Q_i)
\right)
\end{eqnarray}
where $i$ denotes each species of nuclei in the Sun and we will omit
it below, $n$ is the number density of the nuclei, $Q$ is the
energy transfer in the scattering, $Q_{max}(Q_{min})$ denotes the
possible maximum(minimum) energy transfer and $Q_{cap}$ is the
minimal energy transfer which requires DM could be captured only
with a velocity less than $v$, and we define lower limit of
integration as $Q_{min}^{'}=max\left(Q_{cap},Q_{min}\right)$. These
energy transfer functions are given by \cite{Nussinov:2009ft}
\begin{eqnarray}
Q_{max} &=& \frac{1}{2}\mX w^2 \left \{1 - \frac{\mu^2}{\mN^2} \right.
\nonumber \\
&\times & \left. \left(1-\frac{\mN}{\mX}\sqrt{1-\frac{\delta}{\mu w^2/2}} \right)^2 \right \} - \delta\\
Q_{min} &=& \frac{1}{2}\mX w^2 \left \{1 - \frac{\mu^2}{\mN^2} \right.
\nonumber \\
&\times & \left.  \left(1+\frac{\mN}{\mX}\sqrt{1-\frac{\delta}{\mu w^2/2}} \right)^2 \right \} - \delta\\
Q_{cap} &=& \frac{1}{2}\mX (w^2-v^2)  - \delta \ ,
\end{eqnarray}
and one has to impose the condition \bea Q_{max}>Q_{cap} \label{c2}
\eea for the capture to happen. $\sigma^{SI(SD)}$ is the inelastic
SI (SD) cross section and $F(Q)$ is the SI form factor. The term
$\sigma^{SD} S (Q)$ will be decomposed into SD cross sections and SD
form factors for different spin $J$ later. The inelastic SI (SD)
cross section is given by
\begin{eqnarray}
\sigma^{SI} &=& \sqrt{1-\frac{\delta}{\mu w^2/2}}
\left(\frac{\mu}{\mu_{\chi n}} \right)^2 \nonumber \\
&\times& \left(\frac{f_p Z + f_n(A-Z)}{f_p} \right)^2 \sigma_n^{SI}
\end{eqnarray}
where $A$ is the atomic mass number, $Z$ is the charge of the
nucleus, $\mu_{\chi n} \equiv m_{\chi} m_n / (m_\chi + m_n)$ is the
DM-nucleon reduced mass, $f_p(f_n)$ is the SI DM coupling to protons
(neutrons) and we normalize it to $f_p = f_n$, $\sigma_n^{SI}$ is
the DM-nucleon SI scattering cross section. We can see that if
$f_p\simeq f_n $, the DM-nuclei SI cross section is proportional to
$A^2$. Moreover, if $m_{\chi}>>m_{N}$, the reduced mass $\mu~m_N$
will provide another enhancement factor of $A^2$. We can expect
heavy nuclei with large abundance in the Sun, such as Fe, to play a
more important role in the capture process. For the SI form factor,
we use the Helm form factor as, \bea F(Q) = 3 e^{- k^2 s^2/2}
\frac{\sin(kr) - k r \cos(k r)}{(k r)^3} \ , \eea with nuclear skin
thickness $ s = 1$ fm, effective nuclear radius $r = \sqrt{R^2 - 5
s^2}$, $R = 1.2 A^{1/3}$ fm, and momentum transfer $k=\sqrt{2m_N
Q}$.

The SD cross section with form factor for spin-non-zero nuclei is
\begin{eqnarray}
&& \sigma^{SD}S(Q) =\frac{4\pi}{3 a_p^2 (2J+1)}
\sqrt{1-\frac{\delta}{\mu w^2/2}}\left(\frac{\mu}{\mu_{\chi n}}
\right)^2 \nonumber \\
&\times& \left( a_0^2 S_{00}(k)+a_0a_1 S_{01}(k)+a_1^2
S_{11}(k)\right) \sigma_n^{SD} \ ,
\end{eqnarray}
where $a_0=a_p+a_n$ and $a_1=a_p-a_n$ are isoscalar spin and
isovector spin couplings, $a_p$ and $a_n$ are the effective spin
DM-proton and DM-neutron couplings which we assume $a_p=$1 and
$a_n=$0 in the below, $S_{00}(k) , S_{00}(k), S_{01}(k)$ are nuclear
structure functions. In many nuclei models, the nuclear spin is
carried by the unpaired nucleons. Therefore we focus on the nuclei
with an odd nucleon number in the SD scattering process. In the
calculation we use the nuclear structure functions provided by Ref.
\cite{Bednyakov:2006ux}.

\begin{figure}[t]
\includegraphics[totalheight=2.1in]{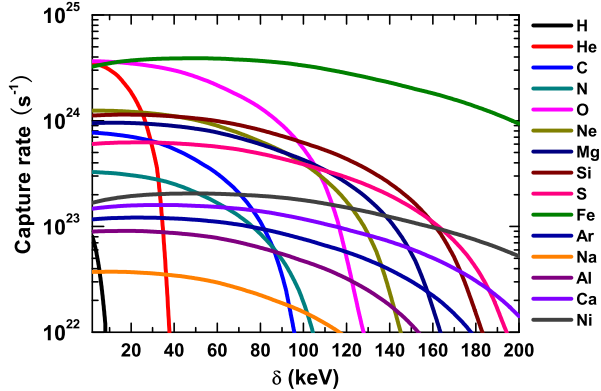}
\caption{Capture rate of 50 GeV spin-independent iDM with
DM-nucleon cross section of $\sigma_n^{SI}\sim$10$^{-4}$pb as a
function of the mass splitting $\delta$ due to different species of
nuclei in the Sun. We choose $\rho_{\chi}=0.3$ GeV cm$^{-3}$,
$\vsun=220$ km/s and $v_0=220$ km/s in the Maxwell-Boltzmann
distribution. }
\label{Fig:cap_iSI_50}
\end{figure}
\begin{figure}[t]
\includegraphics[totalheight=2.1in]{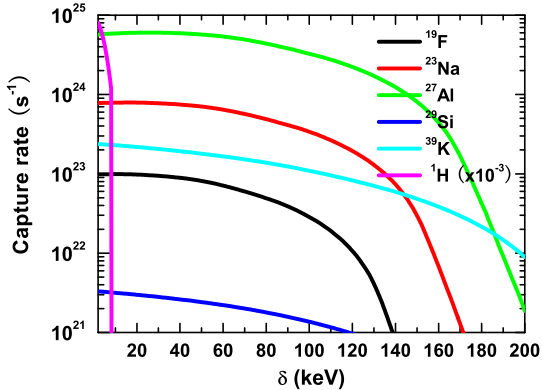}
\caption{Capture rate of 50 GeV spin-dependent iDM with
$\sigma_n^{SD}\sim$ 10 pb as a function of the mass splitting $\delta$
due to different species of nuclei in the Sun.}
\label{Fig:cap_iSD_50}
\end{figure}

The capture rate of 50 GeV DM due to different species of nuclei in
the Sun as a function of the mass splitting $\delta$ is calculated
for both iSI and iSD DM. The results depend on some astrophysical
parameters and as a reference point, we choose $\rho_{\chi}=0.3$ GeV
cm$^{-3}$, $\vsun=220$ km/s and $v_0=220$ km/s in the
Maxwell-Boltzmann distribution. The element abundances are given by
Ref. \cite{Asplund:2009fu} in the form of $\epsilon_i=\log
(n_i/n_H)+12$ and the mass fraction of hydrogen (helium) is 0.7381
(0.2485). In the calculation for SI capture, we pick out the species
with $\epsilon_i > 6$. For SD capture, we calculate $^1$H, $^{19}$F,
$^{23}$Na, $^{27}$Al, $^{29}$Si and $^{39}$K. As we can see in FIG.
\ref{Fig:cap_iSI_50} and \ref{Fig:cap_iSD_50}, Fe and Al provide
most contributions to the iSI and iSD scattering processes
respectively and one can safely neglect the Hydrogen contribution in
the iSD scattering for $\delta > 10$ keV. The capture rate is always
sensitive to the abundance of  heavy nuclei.

Finally, we give some short comments on the possible uncertainty in
the calculation for solar DM capture rate. The input astrophysics
parameters are the local DM mass density $\rho_{\chi}$, velocity of
the Sun $\vsun$, average local circular velocity $v_0$, and solar
elemental abundances $\epsilon_i$. The capture rate is proportional
to $\rho_{\chi}$ and the results will increase if a larger local DM
mass density is available \cite{Bruch:2009rp} or the solar system is
passing through DM substructure \cite{Koushiappas:2009ee}. We also
assume the DM infinity velocity distribution is a Maxwell-Boltzmann
distribution which depends on parameters $\vsun$ and $v_0$. As
discussed in Ref. \cite{Menon:2009qj}, if these two parameters vary
in the region of 200 km/s $< \vsun, v_0 < $ 300 km/s, the capture
rate will be changed by a factor less than two. One should also
notice that, if the velocity distribution deviates from an ordinary
Maxwell-Boltzmann distribution, the result will also be changed. In
our discussions, for simplicity, we assume the solar chemical
composition does not change in the Sun. The capture rate is
sensitive to the abundance of heavy nuclei and increasing the number
density of heavy nuclei in the center of the Sun will increase the
overall capture rate by a few \cite{Menon:2009qj}. There are also
uncertainties arising from the nuclear form factor. For the SI
scattering, the Helm form factor we used in the calculation is less
accurate at large momentum transfers and might be improved by using
some more precise form factors \cite{Duda:2006uk}. For the SD
scattering, the nuclear structure functions in Ref.
\cite{Bednyakov:2006ux} are only considered for direct DM search on
the Earth, so the $q_{max}$ used in the finite momentum transfer
approximation may not be large enough for DM-nucleus scattering in
the Sun if our DM is heavy.

\section{Limits from the Neutrino Telescopes}
\label{sec:limnt}

\begin{figure*}[t]
\includegraphics[totalheight=2.3in,clip=]{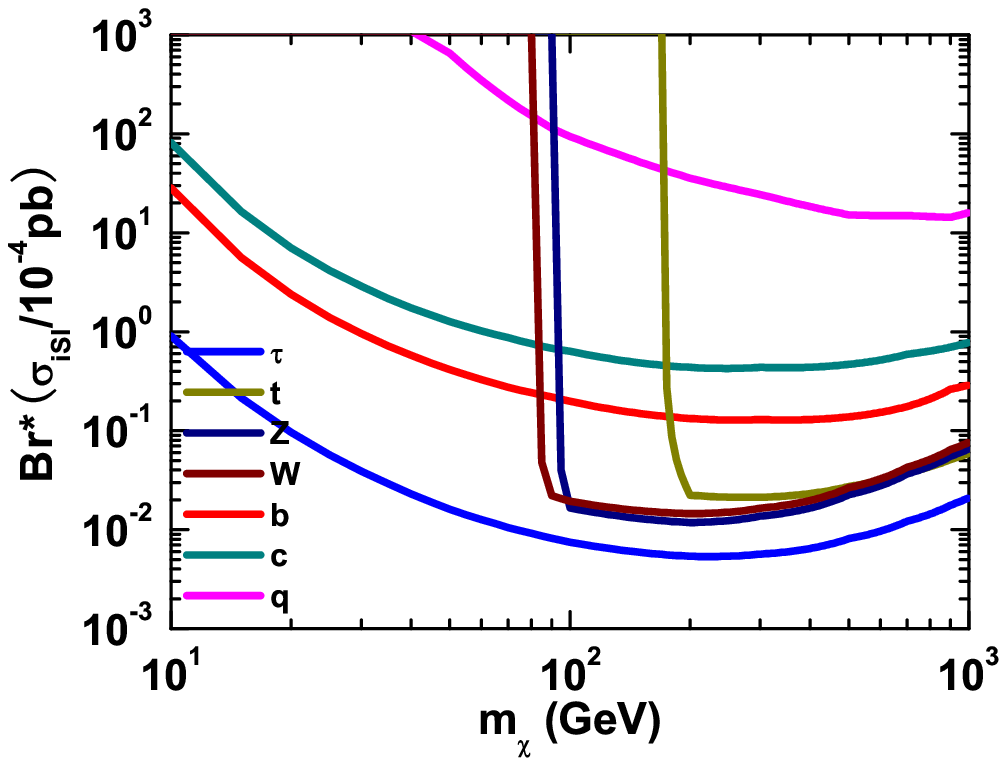}
\includegraphics[totalheight=2.3in,clip=]{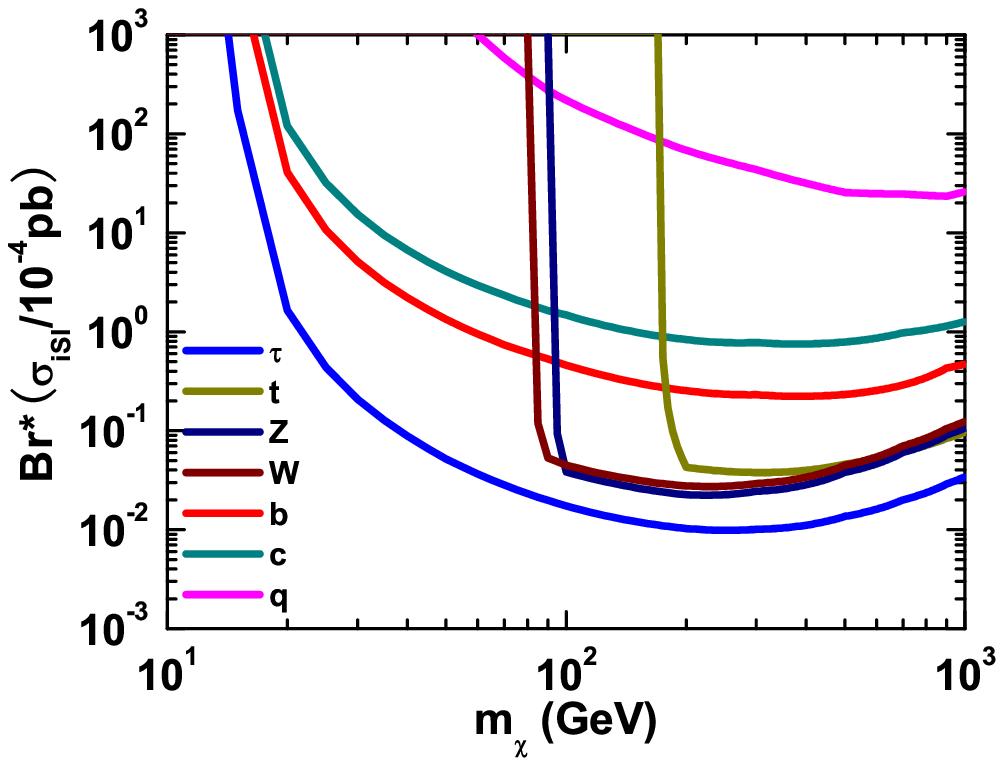}
\caption{Super-K solar neutrino limits on the cross section per
nucleon times the branching ratio to different annihilation channels
for the iSI DM with mass splitting $\delta = 40$ keV (left) and
$\delta = 130$ keV (right). We show the limits on seven annihilation
channels including $\tau \bar{\tau}$, $t \bar{t}$, $W^+W^-$, $ZZ$,
$b \bar{b}$, $c \bar{c}$ and $q \bar{q}$ ($q$ denotes light quark
here). }%
\label{skliSI}%
\end{figure*}
\begin{figure*}[t]
\includegraphics[totalheight=2.3in,clip=]{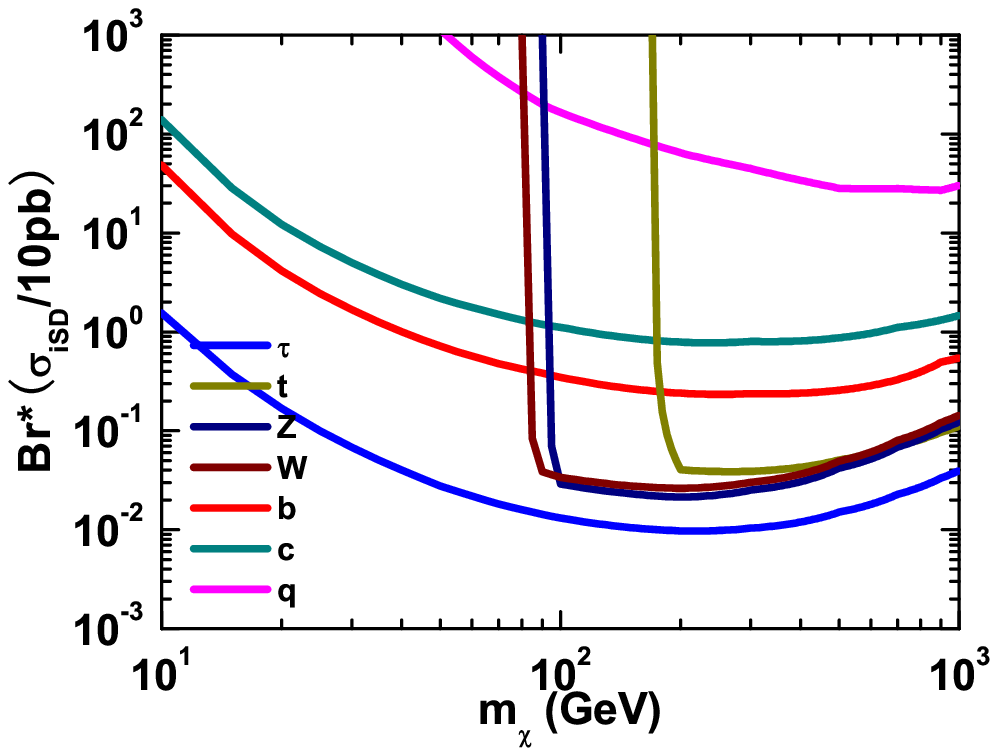}
\includegraphics[totalheight=2.3in,clip=]{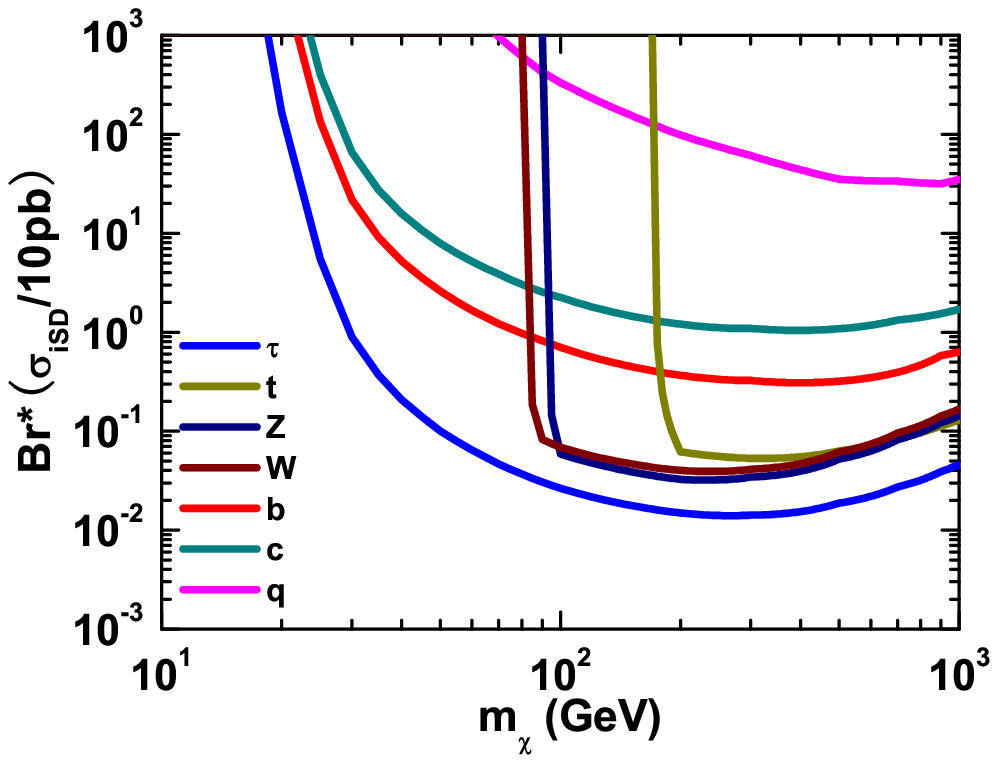}
\caption{Super-K solar neutrino limits on the cross section per nucleon times the branching ratio to different annihilation channels for the iSD DM with mass splitting $\delta = 40$ keV (left) and $\delta = 130$ keV (right).}%
\label{skliSD}%
\end{figure*}
\begin{figure*}[t]
\includegraphics[totalheight=2.3in,clip=]{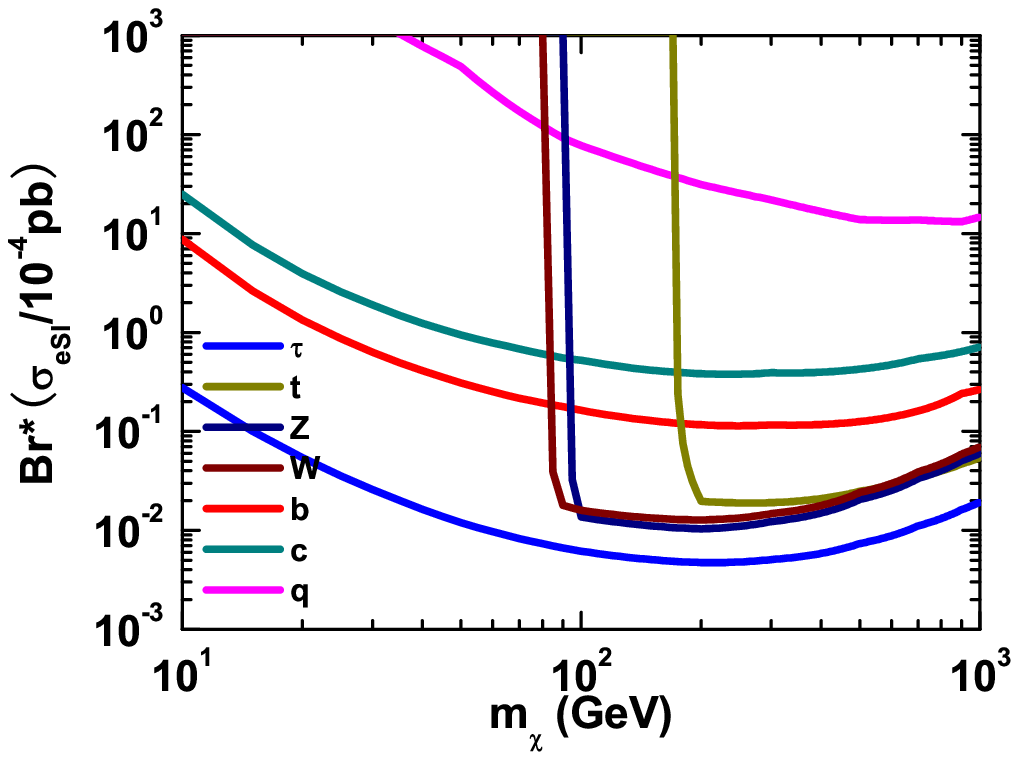}
\includegraphics[totalheight=2.3in,clip=]{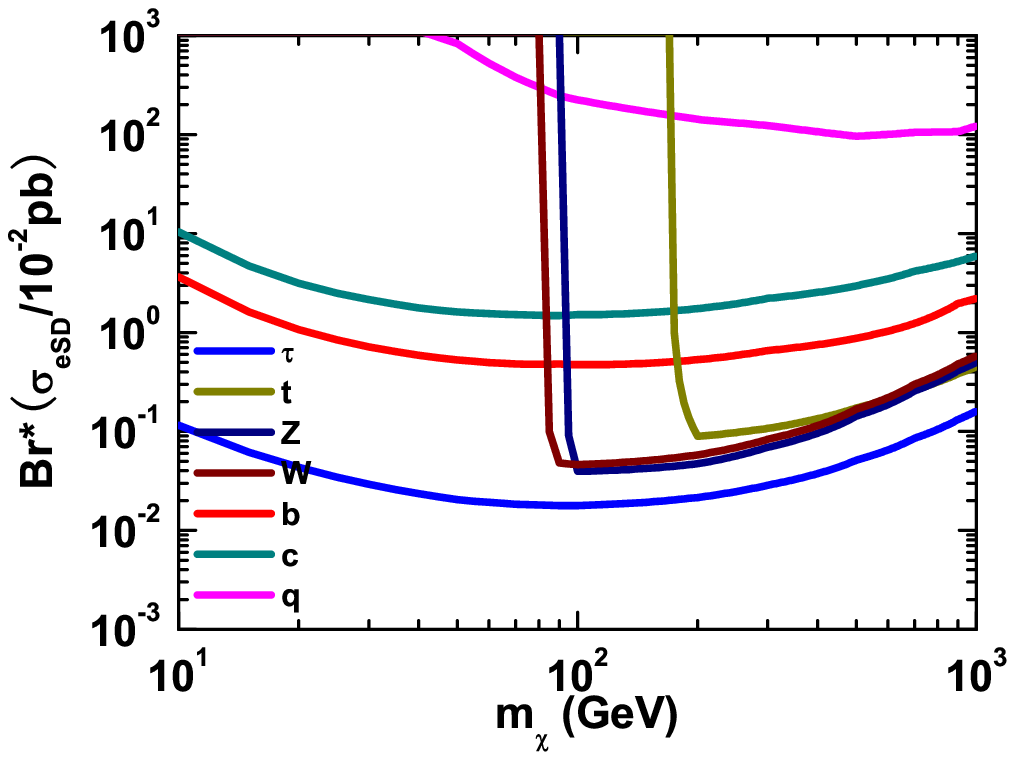}
\caption{Super-K solar neutrino limits on the cross section per nucleon times the branching ratio to different annihilation channels for the elastic spin-independent (eSI) DM (left) and elastic spin-dependent (eSD) DM (right).}%
\label{skle}%
\end{figure*}

If the DM capture and annihilation processes reach equilibrium, we can
simply get the DM annihilation rate
\begin{equation}
\Gamma_A=C_{\odot}/2
\end{equation}

The different DM annihilation channels give different initial
neutrino spectra \cite{Cirelli:2005gh}. If the DM annihilates to
$e^+e^-$ or $\mu^+\mu^-$, they will not contribute to neutrino
signals. This is because for muons, they always lose most of their
energy before decaying in the center of the Sun. For annihilation
channel to $\tau^+\tau^-$, the neutrinos are induced by $\tau$
leptonic decays $\tau\rightarrow \mu \nu \nu$, $e \nu \nu$, and
hadronic decays such as $\tau\rightarrow \pi \nu$, $K \nu$, $\pi \pi
\nu$ et al. For the heavy final annihilation states $W^+W^-$, $ZZ$,
$t \bar{t}$, they produce neutrinos via cascade decays and the
neutrino spectra for such channels are hard. For quark channels, the
hadronization process produces mesons and baryons, and then induces
neutrinos via hadron decays. The neutrino spectra for such channels
are soft. Notice that the light mesons easily lose their energy
before decays, so the contributions to the neutrino signals from
light quarks are usually small.

The neutrinos produced at the solar center will interact with the
matter in the Sun. These effects include the neutral current (NC)
interaction, the charged current (CC) interaction, and tau neutrino
$\nu_{\tau}$ reinjection from secondary tau decays. The
neutrino oscillations including the vacuum mixing and the MSW matter
effects are also important. Here we use the results given by Ref.
\cite{Cirelli:2005gh} to take into account the above effects during the
neutrinos' propagation. The differential neutrino flux at the Earth
is
\begin{equation}
\frac{dN_{\nu}}{dE_{\nu}}=\frac{\Gamma_A}{4\pi R_{SE}^2} \sum_i Br_i
\left(\frac{d N_{\nu}}{d E_{\nu}}\right)_i
\end{equation}
where $i$ runs over the different DM annihilation channels with
branching ratios $Br_i$, $d N_{\nu}/dE_{\nu}$ is the neutrino
spectrum at the Earth after propagation, and $R_{SE}$ is the Sun-Earth
distance.

We can calculate the muon rate at the detector as \footnote{A
different formula that accounts for the energy dependent muon flux
can be found in Ref. \cite{Erkoca:2009by}.}
\begin{eqnarray}
\frac{d\phi_{\mu}}{dE_{\mu}}&=&\int_{E_{th}}^{\mX} dE_{\nu_{\mu}} \;
\frac{dN_{\nu_{\mu}}}{dE_{\nu_{\mu}}}[
\frac{d\sigma_{\text{CC}}^{\nu
p}(E_{\nu_{\mu}},E_{\mu})}{dE_{\mu}}\,n_p \nonumber \\
&& +(p\rightarrow n)]R(E_{\mu}) +(\nu\rightarrow\bar{\nu})
\end{eqnarray}
where $n_p$ ($n_n$) is the number density of protons (neutrons) in
matter around  the detector, the muon range $R(E_{\mu})$ is the
distance that a muon could travel in matter before its energy
drops below the detector's threshold energy $E_{th}$ which is given by
\begin{equation}
R(E_\mu) = \frac{1}{\rho \beta} \ln {\frac{ \alpha + \beta E_{\mu}}{
\alpha + \beta E_{th}} } ,
\end{equation}
with $\alpha , \beta$ the parameters describing the energy loss of
muons in matter.

The cross sections of deep inelastic neutrino-nucleon scattering
processes are given by \cite{Barger:2007xf}
\begin{eqnarray}
&&\frac{d\sigma_{\text{CC/NC}}^{\nu\, p,n}(E_{\nu}, y)}{dy}
\simeq\frac{2\,m_{p,n}\,G_F^2}{\pi}\,E_{\nu}\nonumber \\
&\times&
\left(a_{\text{CC/NC}}^{\nu\,p,n}+b_{\text{CC/NC}}^{\nu\,p,n}\,(1-y)^2\right)
,
\end{eqnarray}
where $y\equiv 1-E_{\ell}/E_{\nu}$. We can see that the neutrino
detector is more powerful to observe neutrino signals in the high
energy region. Because the neutrino-nucleon cross section and muon
range increase as $E_{\nu}$, while the atmospheric neutrino
background decrease as $E_{\nu}^3$.

To obtain limits on the cross section per nucleon times the
branching ratio $\sigma_n \cdot Br_i$, we use the upper bound on
high energy up-going muon flux in the direction of Sun given by
Super-K group \cite{Desai:2004pq}. The Super-K detector is located in a
mine with 1000 m rock overburden which contains 50,000 ton water. The
detector can detect up-going muons with measured path length of at least
7 m which is equivalent to the detector energy threshold 1.6 GeV.
So the Super-K detector has a wide detecting range for high energy
neutrino. Super-K group analyzed the data in 1679.6 days of detector
live time, and no events induced by DM annihilation/decay are confirmed
in the directions of the Sun, the center of Earth, and the Galactic
Center. A stringent limit for up-going muon flux $~O(10^{-15})$
cm$^{-2}$ s$^{-1}$ constrains the DM model. Here we use
the constraint in Ref. \cite{Desai:2004pq} for all annihilation
channels.

The Super-K limits on $\sigma_n \cdot Br_i$ to different
annihilation channels for iSI and iSD DM with different mass
splitting $\delta$ are presented in FIG. \ref{skliSI}, and
\ref{skliSD} respectively. For a typical DM annihilation cross
section of $10^{-4}$ pb (or higher) that explains DAMA in the iSI DM
case, the limits are quite constraining on hard channels ($W^+W^-$,
$ZZ$, $t\bar{t}$, $\tau^+ \tau^-$), are lessened in softer channels
like $c \bar{c}$ and $b \bar{b}$ and are quite loose in the light
quark channels. In the iSD DM case, the typical DM annihilation
cross section (or higher) to explain DAMA is enhanced to $10$ pb
\cite{Kopp:2009qt}. Interestingly, even though the cross section is
$10^5$ times higher than the iSI DM case, we reach a similar
conclusion that the limits are quite constraining on hard channels,
are lessened in softer channels like $c \bar{c}$ and $b \bar{b}$ and
are quite loose in the light quark channels. In order to see how the
inelastic property of DM nucleus scattering changes the limits, we
also calculate the Super-K limits on $\sigma_n \cdot Br_i$ to
different annihilation channels for eDM. In the SI case, making the
scattering inelastic changes the results at most an order of
magnitude for reasonable $\delta$. In the SD case, however, the
limit is loosened by 3 orders of magnitude when hydrogen no longer
contributes to the dark matter capture in the Sun.

The on-going large volume neutrino telescopes, such as IceCube and
KM3NET, are more powerful to probe high energy neutrinos. IceCube
located between depth of 1.45 km and 2.45 km in the south pole will
have an effective detecting area of 1 km$^2$. In 2007, IceCube
consisted of 22 strings has provided a more stringent upper limit to
muon flux from the Sun \cite{Abbasi:2009uz}. The main shortcoming of
IceCube 22-strings is its high energy threshold. So the experimental
results are only useful to constrain the heavy DM \footnote{For the
$b\bar{b}$ and $W^+W^-$ channels, the constraints are available for
DM heavier than 250 GeV and 500 GeV respectively. For the
constraints to iDM, one can see the Ref. \cite{Menon:2009qj}.}. The
complete detector consisted of 80 strings might reduce the energy
threshold to 50 GeV. We need to notice that the effective detecting
area of IceCube decreases rapidly in the low energy region. In
addition, the angular resolution becomes worse in such energy
region. So the capability of IceCube to search for light DM is
weaker than that for heavy DM.

\begin{figure*}[t]
\includegraphics[totalheight=2.3in,clip=]{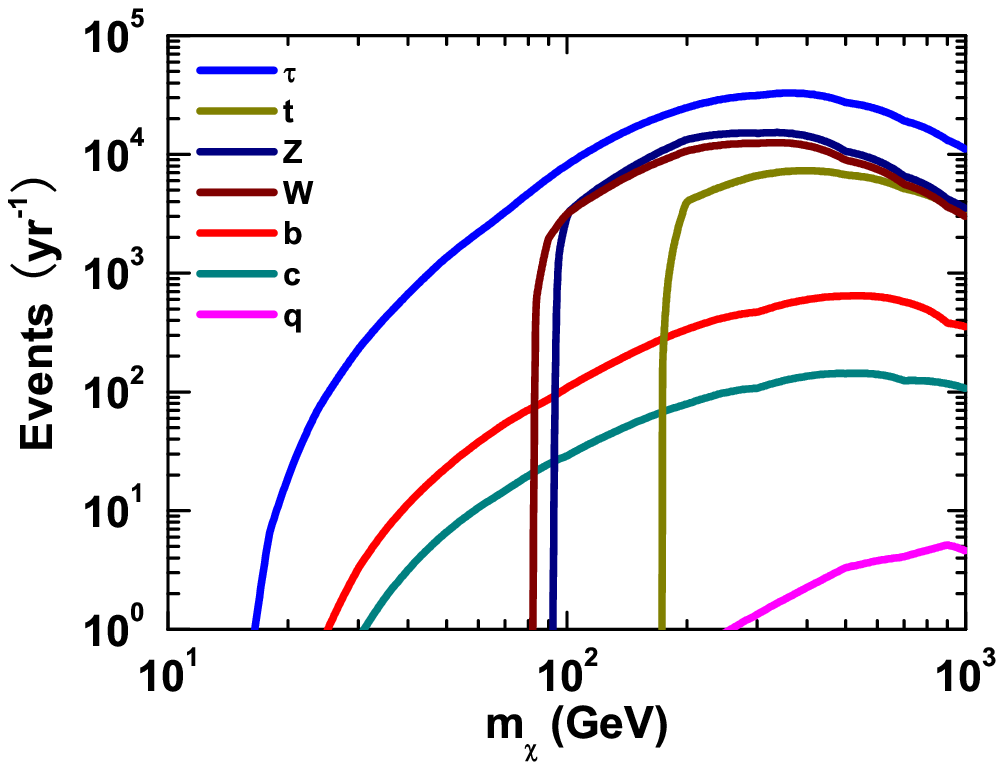}
\includegraphics[totalheight=2.3in,clip=]{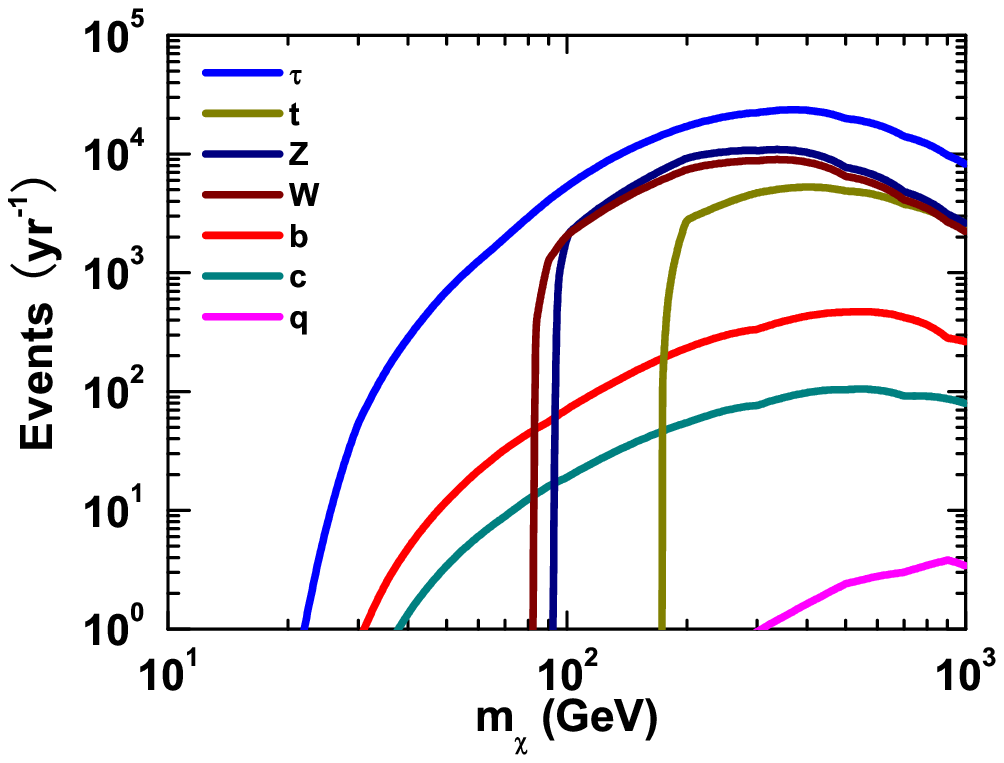}
\caption{(left) Neutrino event number in one year at IceCube (with
DeepCore) for the iSI DM with $\sigma_n^{SI}=$ 10$^{-4}$ pb
(left) and iSD DM with $\sigma_n^{SD}=$ 10 pb (right). We
choose mass splitting
$\delta = 130$ GeV here.}%
\label{icemunum}%
\end{figure*}

But it is worth remarking, additional IceCube 6 special strings
named DeepCore, are powerful to improve the capability of the
detector and reduce the detection threshold down to 10 GeV
\cite{Resconi:2008fe, Wiebusch:2009jf}. In the FIG. \ref{icemunum},
we show the total number of events at the IceCube plus DeepCore.
Here we use the effective neutrino detection area of IceCube plus
DeepCore given by Ref. \cite{Wiebusch:2009jf} to calculate the event
rate at the detector. The number of background events from
atmospheric neutrinos within a 3$\degree$ window is estimated
$\sim$50 in the energy range $E \sim (10, 200)$ GeV. For heavy iDM
$m_{\chi}
>$100 GeV, the IceCube plus DeepCore is very powerful to test most
of the iDM annihilation channels, which include all hard channels
and soft channels to $b \bar{b}$ and $c \bar{c}$. For the iDM
lighter than 100 GeV, comparing with FIG. \ref{skliSI} and
\ref{skliSD} based on the Super-K 90$\%$ C.L. exclusion limit, we
can see that the IceCube plus DeepCore also has capability to test
or further constrain the iDM model with soft annihilation channels
to $b \bar{b}$ and $c \bar{c}$.

\section{Neutrino Constraint from Terrestrial Inelastic Dark Matter Annihilations}

\begin{figure*}[t]
\includegraphics[totalheight=2.3in,clip=]{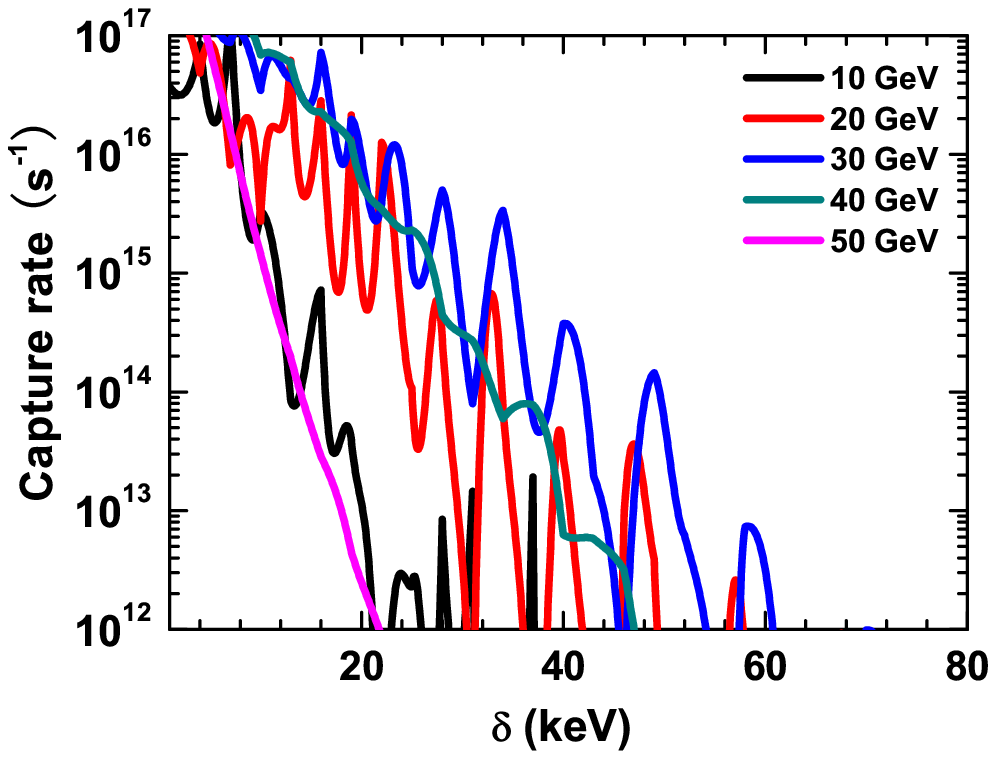}
\includegraphics[totalheight=2.3in,clip=]{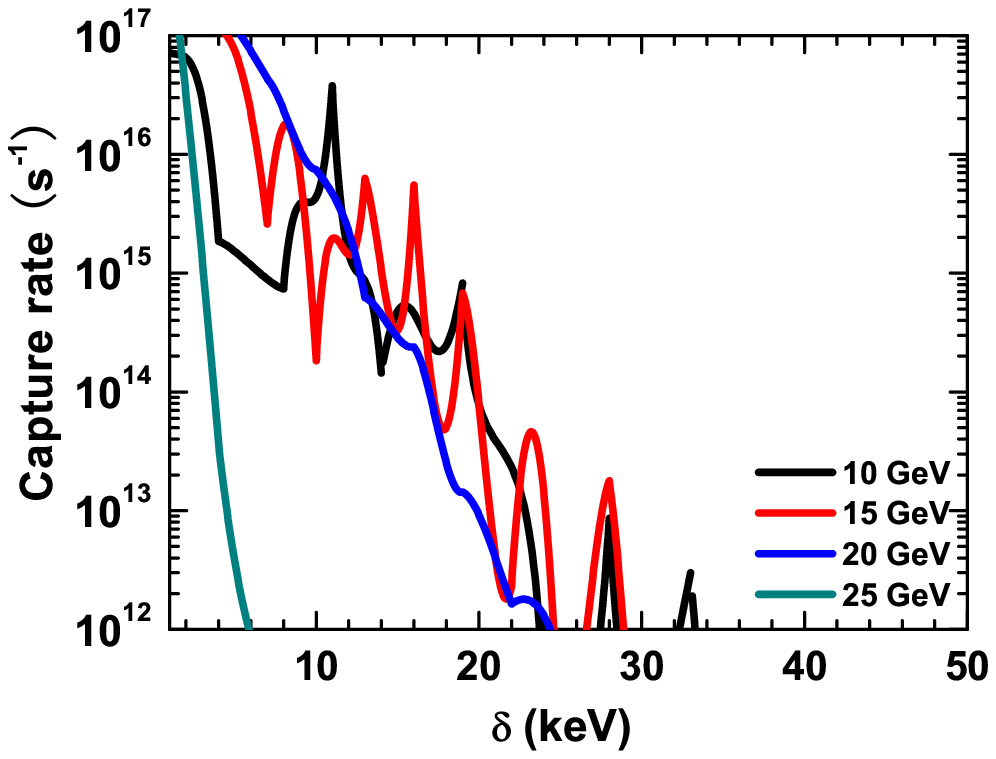}
\caption{Terrestrial iDM capture rate for the iSI DM with $\sigma_n^{SI}=$ 10$^{-4}$ pb (left) and iSD DM with $\sigma_n^{SD}= $10 pb (right).}%
\label{capea}%
\end{figure*}
\begin{figure*}[t]
\includegraphics[totalheight=2.3in,clip=]{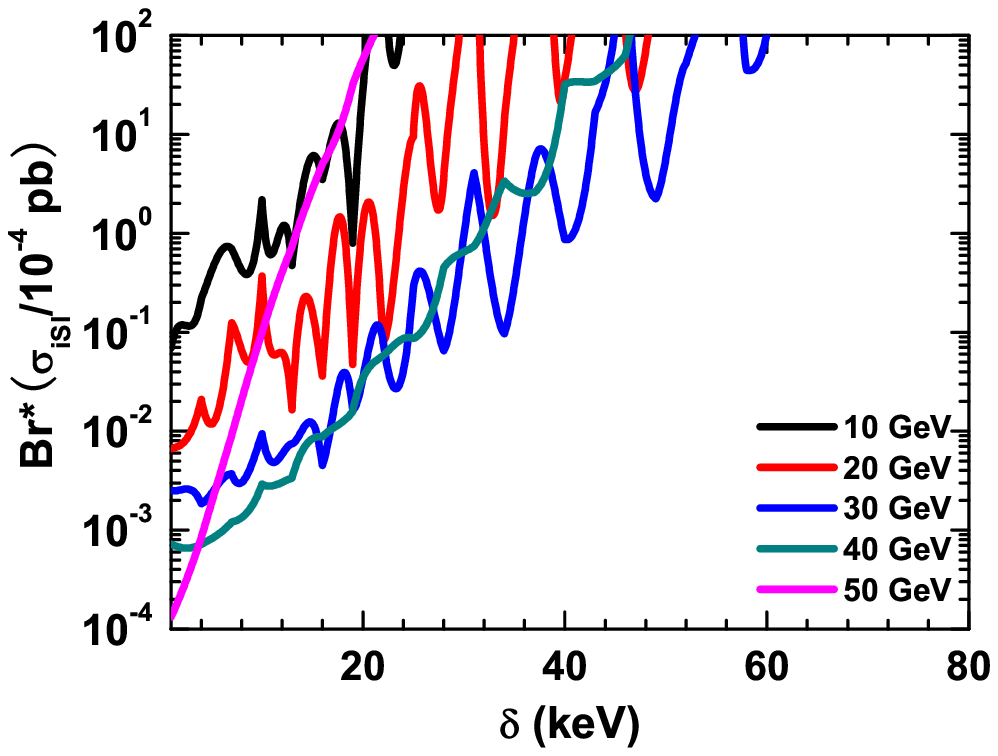}
\includegraphics[totalheight=2.3in,clip=]{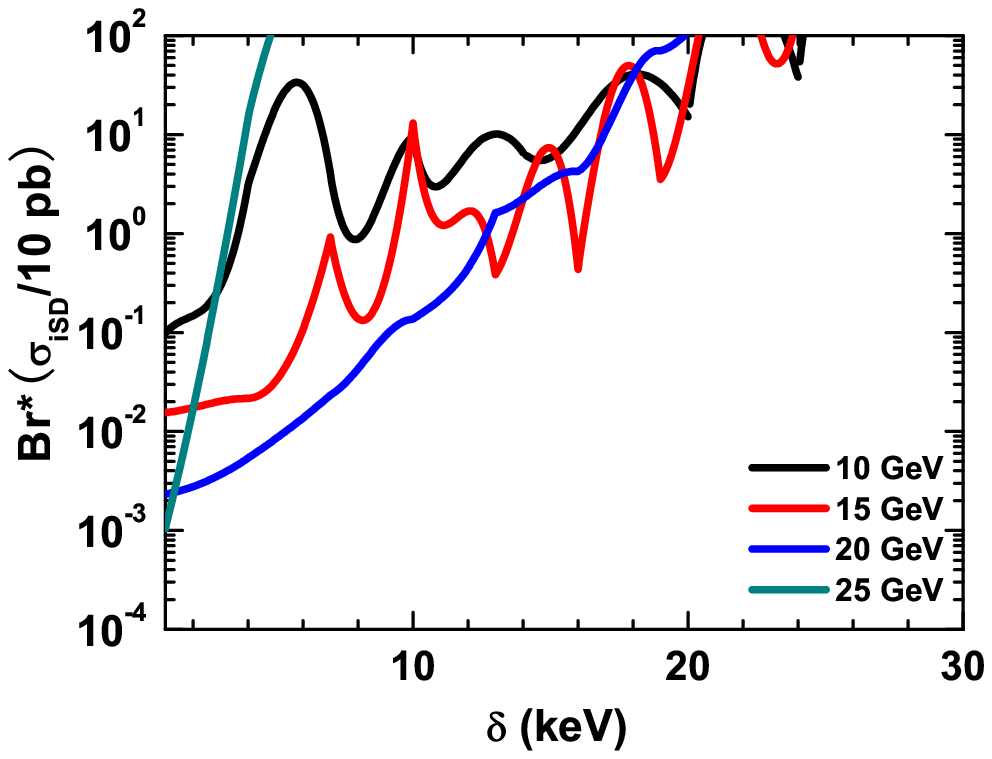}
\caption{Super-K terrestrial neutrino limits on the cross section per nucleon times the branching ratio to different annihilation channels for the iSI DM (left) and iSD DM (right). }%
\label{sklea}%
\end{figure*}

\label{sec:limearth} Unlike the Sun, the main constituents of the
Earth are heavy nuclei, such as O, Al, Si, Mg, Fe and Ni. Since the
abundances of Fe and Al are 31.9\% and 1.59\% respectively, we
expect that the abundance of such heavy elements helps the capture
process. Nevertheless, probing neutrino signals from terrestrial iDM
annihilations is more difficult than probing them from solar iDM
annihilations. There are two main reasons for this. First, as a
gravitational system, the Earth is far smaller than the Sun with a
rather small escape velocity, so it is more difficult to trap DM in
the center of the Earth. In the iDM scenario, the situation becomes
even worse due to the kinetic suppression induced by inelastic
capture condition. Second, it is well known that terrestrial DM
capture and annihilation processes can not reach equilibrium in the
ordinary DM scenario, so the annihilation rate of DM trapped in the
Earth is even smaller than half of the capture rate.

Now we turn to Eq. (\ref{c1}) and (\ref{c2}) and see the first
reason of suppression explicitly. We can rewrite Eq. (\ref{c2}) as
\begin{equation}
\frac{\chi'}{\chi_{-}^{'2}}>\frac{u^2}{v^2} \label{c3}
\end{equation}
where we use a similar notation as Ref. \cite{Gould:1987ir}
\begin{equation}
\chi\equiv \frac{m_{\chi}}{m_N} \;\; , \; \; s\equiv
\sqrt{1-2\delta/(\mu w^2)},
\end{equation}
and
\begin{equation}
\chi'_{-}\equiv\frac{\chi-s}{2} \;\; , \;\; \chi'\equiv \chi
\left (\frac{1+s}{2} \right )+\frac{1-s^2}{4} \ .
\end{equation}
The escape velocity of the Earth $v\simeq$ 10 km/s is very small
compared to the escape velocity of the Sun $\simeq$ 10$^3$ km/s and
the velocity dispersion of local DM $u \sim \bar{v}\simeq$ 270 km/s,
so the right hand side of the capture condition Eq. (\ref{c3}) is a
large value. When the mass of DM is comparable to or larger than
that of a nucleus, such as Fe or Al, the left hand side of Eq.
(\ref{c3}) is suppressed by $m_{\chi}$ so it is difficult to satisfy
the capture condition. If the DM is much lighter than the nuclei,
the condition Eq. (\ref{c3}) is still not easy to satisfy. Moreover
it is difficult to satisfy the inelastic scattering condition Eq.
(\ref{c1}) simultaneously. So the most promising situation is that
DM and nuclei have similar masses and $\delta$ must not be large.

The terrestrial iDM capture and annihilation do not reach
equilibrium in ordinary DM scenario \footnote{If the DM annihilation
cross section is far larger than ordinary value 3$\times$10$^{-26}$
cm$^3$/s due to some enhancements such as Sommerfeld effect, the
terrestrial DM capture and annihilation could reach equilibrium
\cite{Delaunay:2008pc}.}. Following from Eq. (\ref{eq:CArate}), the DM annihilation rate is
\begin{equation}
\Gamma_A=\frac{C_{\oplus}}{2} \tanh^2(t_\oplus/t_{eq}),
\end{equation}
where $t_\oplus$ is the age of Earth, $C_\oplus$ is the capture rate
of the DMs in the Earth. The equilibrium time, effective
annihilation rate is defined as $t_{eq} \equiv 1/\sqrt{C_\oplus
C_A}$, $C_A \equiv \langle \sigma v \rangle /V_{\oplus eff}$
respectively. It is also useful to define a critical capture rate
for the Earth as $C_{\oplus}^c \equiv 1/C_A t_{\oplus}^2 \sim
10^{14}s^{-1}(\textrm{TeV}/m_{\chi})^{3/2}$. If $C_\oplus >
C_\oplus^c$, the DM capture and annihilation in the Earth could
reach equilibrium.

It is straightforward to get numerical results by using the method
in Sec II. Here we use the mass density profile and the composition
of the Earth provided by Ref. \cite{Gandhi:1995tf} and
\cite{compoea} respectively. For simplicity, we assume the Earth is
in free space \cite{Gould:1987ir}. If the gravitational interaction
of the Sun is taken into account, there is a correction factor of
O(1) \cite{Gould:1987ww} to the capture rate. In FIG. \ref{capea},
we show the results of terrestrial iDM capture rate for SI and SD
scattering processes. We find that for very heavy or light DM, the
capture process is highly suppressed. So we only give the capture
rate for the iDM with typical mass in the range of (10, 50) GeV
which is similar to the mass of nuclei of interest. It is
interesting to notice that there are surges as $\delta$ is varying.
The peaks of these surges ($\chi'_- \rightarrow 0$) are quite
similar to the ``resonant enhancement" in the eDM case when DM mass
reaches to the nuclei's mass as discussed in Ref.
\cite{Gould:1987ir} and is absent if the DM is not light enough. For
the iDM scenario, we need to notice that for the n species of nuclei
of interest here, there are 2n inelastic capture conditions to be
satisfied or obeyed. So if the mass of DM is similar to that of the
nuclei, the terrestrial capture rate is very sensitive to the
parameters of $m_{N}$, $m_{\chi}$ and $\delta$.

Here we use the upper limit for up-going muon flux from
the center of the Earth provided by Super-K \cite{Desai:2004pq} to constrain the iDM model. We
only show the most stringent constraints assuming all the DM
annihilation products are $\tau^+ \tau^-$. We can see the results in
FIG. \ref{sklea} and we find that the constraints from Earth-born
neutrinos are weaker than those from the Sun. The most stringent
constraints arise from small $\delta$ and $m_\chi
\sim$ (10, 50) GeV.

\section{Conclusion and Discussion}

Inelastic dark matter (iDM) is an excellent candidate to explain the
DAMA annual modulation result. Recently, the CDMS II announced the
observation of two signal events and set a more stringent constraint
to the iDM model to account for DAMA results. The mass of the iDM is
not heavier than 100 GeV and the DM-nucleon cross section is
typically 10$^{-4}$ pb (or higher) for iSI case and is 10 pb (or
higher) for iSD case.

In this work, we discuss the constraints by neutrino signals from
solar and terrestrial iDM annihilations motivated after DAMA and
CDMS II results. The neutrino flux from the Sun or Earth depends on
the capture rate and the constraint on DM-nucleon cross section is
quite different from the eDM case. For iSI/iSD scattering, the main
contribution to the capture rate in the Sun is from Fe/Al and in the
latter case, the capture from H is highly kinematically suppressed.
Although the types of nuclei that contribute to the capture rate are
quite different in iSI and iSD DM models, the constraints for the
iDM annihilation channels for the typical DM-nucleon cross section
are actually similar. The Super-K null results set very stringent
constraints to the hard DM annihilation channels such as $W^+W^-$,
$ZZ$, $t\bar{t}$ and $\tau^+ \tau^-$. For the soft channels such as
$b \bar{b}$ and $c \bar{c}$, the limits are loose. With the IceCube
80-strings plus DeepCore, it can discover or exclude such channels.
If the DM annihilation products are light charged leptons $e^+e^-$,
$\mu^+ \mu^-$, light mesons or some new light bosons, there are no
constraints from the neutrino detection. For neutrinos from DM
annihilation in the Earth, the constraint is weaker than the one in
the Sun. This is mainly due to the fact that Earth being a small
gravitational system that captures the DM. Moreover, the terrestrial
iDM capture rate is suppressed due to the inelastic capture
condition. Furthermore, the terrestrial iDM capture and annihilation
do not reach equilibrium in most of its parameter space. The
constraint is available only for small $\delta$ and iDM with mass of
several tens of GeV where similar behaviors as the ``resonance
enhancement" in the eDM case occur.

If the DAMA and CDMS II results are really induced by DM, we could
expect the XENON 100 \cite{Aprile:2009yh} or other direct detection
experiments to give the similar observations in the immediate
future. Nevertheless, the constraints and future detection prospects
we have obtained from neutrino telescope should be considered in any
explanations based on iDM. Other on-going indirect detection
experiments and the Large Hadron Collider also have the capability
to detect the DM signals. It is exciting and useful to think about
all these experiments and achieve a complete picture of DM.

\label{sec:con}

\section{Acknowledgement}

We would like to thank Paul H. Frampton, Sourav Mandal and Matthew Sudano for useful discussions. The work of J.S. was supported by the World Premier International Research Center Initiative (WPI initiative) MEXT, Japan and
Grant-in-Aid for scientific research (Young Scientists (B) 21740169)
from Japan Society for Promotion of Science (JSPS). The work of
P.F.Y. and S.H.Z was supported by the Natural Sciences Foundation of
China (Nos. 10775001, 10635030). P.F.Y. would like to acknowledge the hospitality of
Institute for Physics and Mathematics of the Universe (IPMU) while the work was initiated.

\end{document}